\documentclass[aps,prd,twocolumn,amsmath,groupedaddress]{revtex4}
\usepackage{epsfig}

\def\frac#1#2{\textstyle{{{#1} \over {#2}}}}

\def\lsim{\mathrel{\rlap{\lower4pt\hbox{\hskip1pt$\sim$}}
    \raise1pt\hbox{$<$}}}
\def\gsim{\mathrel{\rlap{\lower4pt\hbox{\hskip1pt$\sim$}}
    \raise1pt\hbox{$>$}}}

\newcommand{\beq}{\begin{equation}}
\newcommand{\eeq}{\end{equation}}
\newcommand{\bea}{\begin{eqnarray}}
\newcommand{\eea}{\end{eqnarray}}
\newcommand{\bse}{\begin{subequations}}
\newcommand{\ese}{\end{subequations}}

\def\sqr#1#2{{\vcenter{\vbox{\hrule height.#2pt
         \hbox{\vrule width.#2pt height#1pt \kern#1pt
         \vrule width.#2pt}
         \hrule height.#2pt}}}}

\begin{document}

%\title{ Superdeformed Oblate Minima in Superheavy Nuclei.}
\title{Superdeformed Oblate Superheavy Nuclei?}
\author{P.~Jachimowicz$^{1,2}$, M.~Kowal$^{1}$, J. Skalski$^{1}$}

\affiliation{$^1$ Soltan Institute for Nuclear Studies, Ho\.za 69,
PL-00-681 Warsaw, Poland} \affiliation{$^2$ Institute of Physics,
University of Zielona G\'{o}ra, Szafrana 4a, 65516 Zielona
G\'{o}ra, Poland}
%\affiliation{$^3$ Gesellschaft f{\"u}r
%Schwerionenforschung (GSI), Planckstrasse 1, D-64291 Darmstadt,
%Germany}

\date{\today}

\begin{abstract}
{\noindent We study stability of superdeformed oblate (SDO) superheavy 
 $Z\geq 120$ nuclei predicted by systematic 
    macroscopic-microscopic calculations in 12D deformation space 
  and confirmed by the Hartree-Fock calculations with the realistic SLy6 force.
  We include into consideration high-$K$ isomers that very likely form at 
  the SDO shape. Although half-lives $T_{1/2}\lesssim10^{-5}$ s
 are calclulated or estimated for even-even spin zero systems, 
 decay hindrances known for high-$K$ isomers suggest that some SDO superheavy 
 nuclei may be detectable by the present experimental technique.  
   }
\end{abstract}
\maketitle

 The question of what is the largest possible atomic number $Z_{max}$ of  
  an atomic nucleus is still unsettled.
 The recent experiments on heavy ion fusion in Dubna claim $Z_{max}\geq 118$
\cite{118}, with a partial confirmation of hot fusion cross-sections coming 
 from GSI \cite{GSI} and LBL Berkeley \cite{LBL}.
  Predictions on the stability of superheavy nuclei are based either on 
  the Hartree-Fock (HF) studies with some effective 
  interaction chosen out of the existing multitude, or on the more 
  phenomenological, but also more tested, macroscopic-microscopic method. 
  Although these models differ quantitatively, they consistently predict 
  prolate deformed superheavy nuclei with $Z=100-112$, which is confirmed 
  experimentally for nuclei around $^{254}$No \cite{No}, and 
 spherical or oblate deformed systems with $Z\geq 114$ and $N=174$-184, see 
 e.g. \cite{611,CHN}. In the present letter we show   
   that realistic calculations predict superdeformed oblate (SDO) nuclei, 
 with characteristic quadrupole deformations $-0.4\lesssim \beta_{20} 
 \lesssim -0.5$ (spheroids with the axis ratio $\approx$ 3:2), 
 for $Z\geq 120$. By this we confirm and extend one of the 
  conflicting conclusions of \cite{611} (Fig.12 there). 
  Relying on the calculated energy surfaces, masses and cranking mass 
  parameters, we calculate or estimate half-lives for selected even-even SDO 
  systems. Then we consider an idea, advanced e.g. in \cite{marinov}, 
  of extra stable high-$K$ shape isomers, also in odd systems, whose 
  existence at the SDO shape is very likely. Expected decay hindrances 
   point to the possibility that some of these exotic-shaped superheavy nuclei, 
  far from the conventionally expected "island of stability", 
  live long enough to be detected.

  {\it The Model}.
 Within the macroscopic-microscopic method, energy of a deformed nucleus is
 calculated as a sum of two parts: the macroscopic one being a smooth
 function of $Z$, $N$ and deformation, and the fluctuating microscopic one
  that is based on some phenomenological single-particle (s.p.) potential.
  A deformed Woods-Saxon potential model used here 
 is defined in terms of the nuclear surface, as exposed in \cite{WS}. 
  We admit shapes defined by the following equation of the nuclear
surface: 
  \bea
  \label{shape}
    R(\theta,\varphi)&=& c(\{\beta\}) R_0 \{ 1+\sum_{\lambda>1}\beta_{\lambda 0}
   Y_{\lambda 0}(\theta,\varphi)+
\nonumber \\
&&
    \sum_{\lambda>1, \mu>0, even}
   \beta_{\lambda \mu c} Y^c_{\lambda \mu} (\theta,\varphi)\}  ,
\eea
  where $c(\{\beta\})$ is the volume-fixing factor. The real-valued spherical
  harmonics $Y^c_{\lambda \mu}$, with even
  $\mu>0$, are defined in terms of the
  usual ones as: $Y^c_{\lambda \mu}=(Y_{\lambda \mu}+Y_{\lambda -\mu})/
 \sqrt{2}$.
  In other words, we consider shapes with two symmetry planes. 
% we took the
% "universal set" of potential parameters and the pairing strengths
%  $G_n=(17.67-13.11\cdot I)/A$ for neutrons, $G_p=(13.40+44.89\cdot I)/A$
%  for protons ($I=(N-Z)/A$).
%The $n_{p}=450$ lowest proton levels and $n_{n}=550$ lowest
%neutron levels from $N_{max}=19$ lowest shell of the oscillator
%are taken into account in the diagonalization procedure.
% We have determined the single - particle spectra for every investigated
%nucleus.
% These calculation not include any scaling relation to the
%\emph{central} nucleus.
%  The Strutinsky smoothing was performed with the
%  6-th order polynomial and the smoothing parameter equal to
%   $1.2 \hbar\omega_0$.
  For the macroscopic part we used the Yukawa plus exponential model
  \cite{KN}. 
%  Neither symetry energy nor Wigner term in macroscopic part is used.
  All parameters used in the present work, determining 
   the s.p. potential, the pairing strength and the 
  macroscopic energy, are equal to those used previously in the calculations 
  of masses \cite{WSpar} and fission barriers \cite{Kow} of heaviest 
  nuclei.

  {\it Calculations.}
 We used a rich variety of shapes, with possible nonaxiality and
 mass-asymmetry, to reliably determine energy landscapes of the heaviest
 nuclei.
  A deformation set included both traditional
   quadrupole deformations $\beta$ and $\gamma$, where
   $\beta_{20}=\beta\cos\gamma$ and $\beta_{22c}=-\beta\sin\gamma$
    (for $\gamma=n\times 60^o$, with $n$ integer, a quadrupole shape is
     axially symmetric), three hexadecapole
   distortions $\beta_{40}, \beta_{42c}, \beta_{44c}$, the higher-rank even
  axial multipoles $\beta_{60}$ and $\beta_{80}$, and the following
  odd-multipole deformations: $\beta_{30}, \beta_{32c}, \beta_{50},
  \beta_{52c}$ and $\beta_{70}$ - altogether twelve parameters.
  The range of deformation parameters covered a region of shapes up to,
  and little behind the fission barrier, where the shape parametrization
  (\ref{shape}) may be hoped sufficient.

 Energy landscapes were obtained by a multidimensional energy
  minimization on a map of equidistant mesh points ($\beta\cos\gamma,
  \beta\sin\gamma$) with respect to 10 other deformations.
   We used a rather large mesh spacing of 0.05 in order to make time-consuming
   calculations feasible. A subsequent interpolation served to visualize
   results. In order to check the results we monitored the
   continuity of the resulting 10 deformation parameters with respect
   to $\beta\cos\gamma$ and $\beta\sin\gamma$, and their stability
   with respect to the choice of their starting values.
  To assess the latter, we repeated the minimization for
  the whole map for selected nuclei by choosing random starting values.
   We have found that the results agreed with the ones obtained previously.
%  The starting values of $\beta_{30}$
%  and $\beta_{32c}$ were always taken different from zero.
%  Still, it should be noted that we cannot be absolutely certain that the
%  found 10D minima are exclusively global.
    Additional minimizations have been done to further verify the found 
    minima, in particular, the axially symmetric minima were reproduced 
    by the minimization over the axially symmetric deformations  
     $\beta_{\lambda 0}$.

\begin{figure}[t]
\vspace{-5mm}
\hspace{-3mm}
\begin{minipage}[l]{-2mm}
%\centerline{\includegraphics[scale=0.85]{b2.PDF}}
\centerline{\includegraphics[scale=0.85]{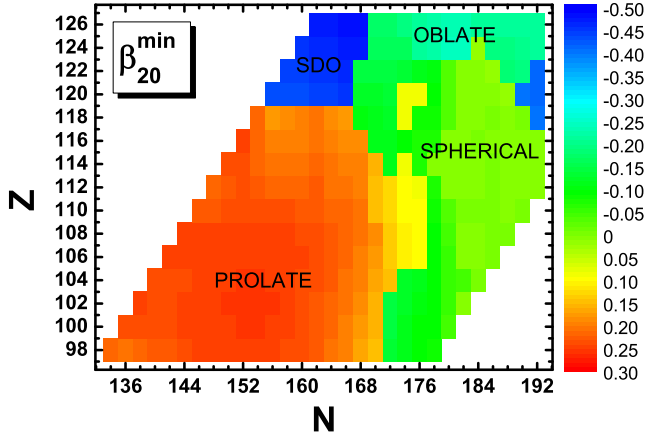}}
\end{minipage}
\caption{{\protect Calculated ground state quadrupole deformations $\beta_{20}$
   (color online).  }}
  \label{fig:b2}
\end{figure}

\begin{figure}[h]
\vspace{-5mm}
\begin{minipage}[h]{35mm}
\centerline{\includegraphics[scale=0.42]{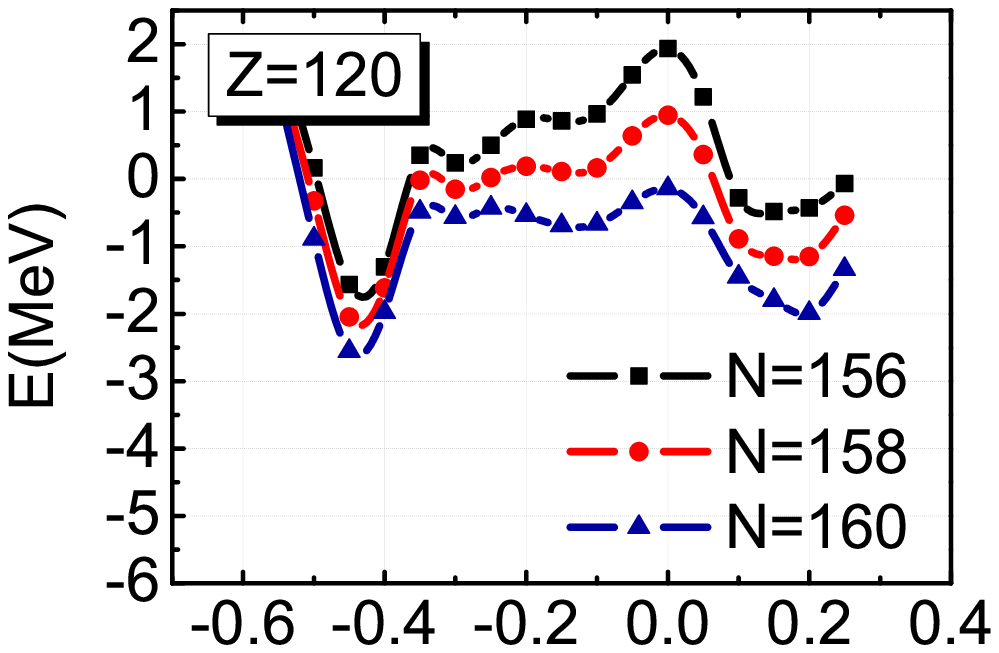}}
\vspace{-7mm}
\centerline{\includegraphics[scale=0.42]{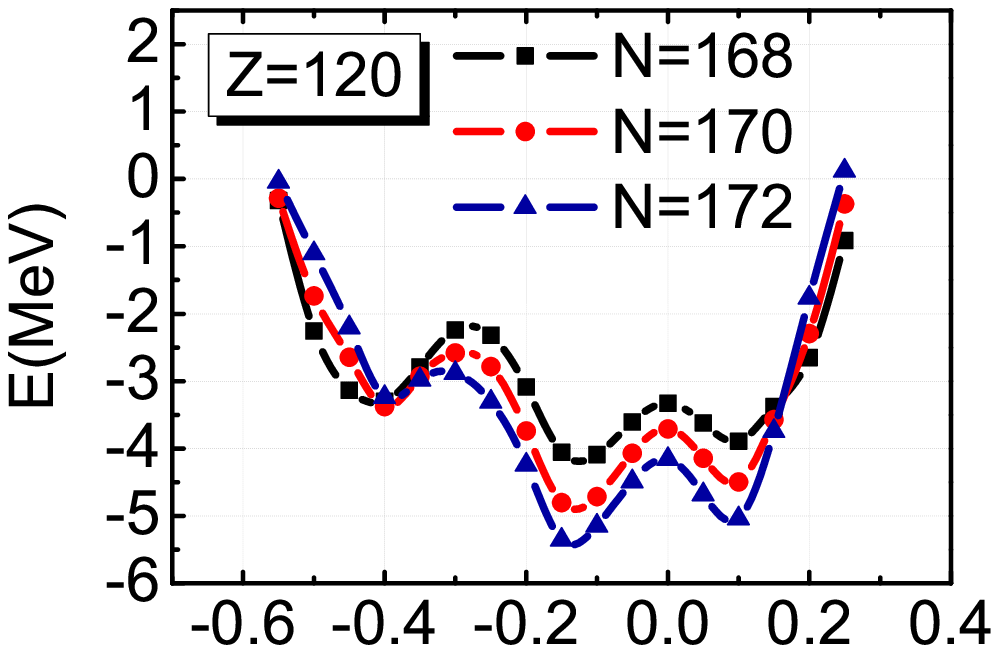}}
\vspace{-7mm}
\centerline{\includegraphics[scale=0.42]{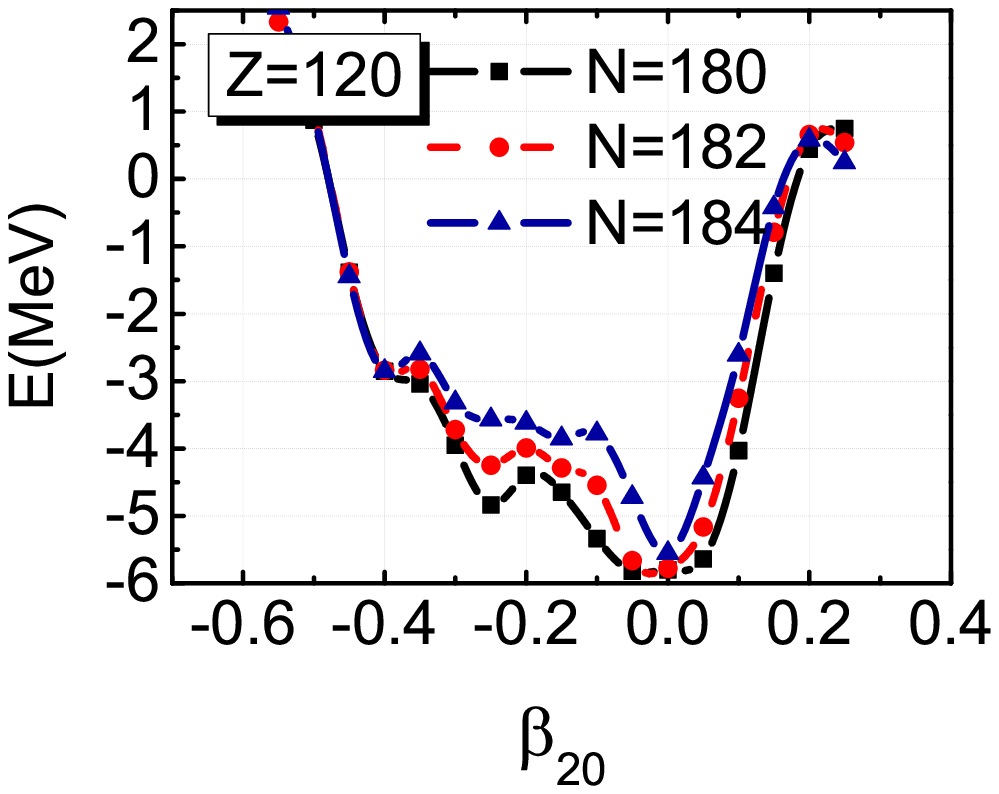}}
\end{minipage}
\begin{minipage}[h]{+50mm}
\centerline{\includegraphics[scale=0.42]{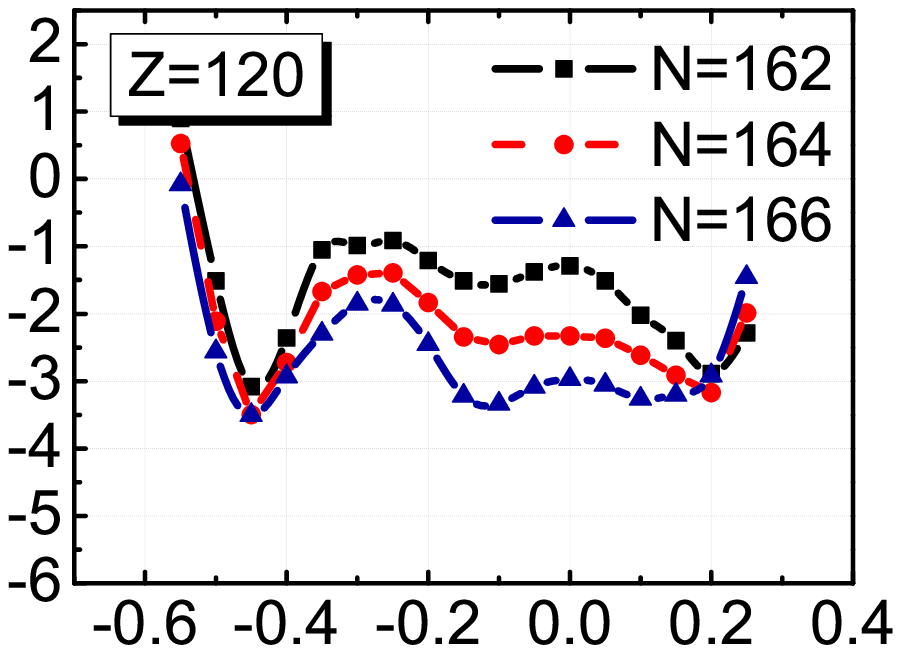}}
\vspace{-7mm}
\centerline{\includegraphics[scale=0.42]{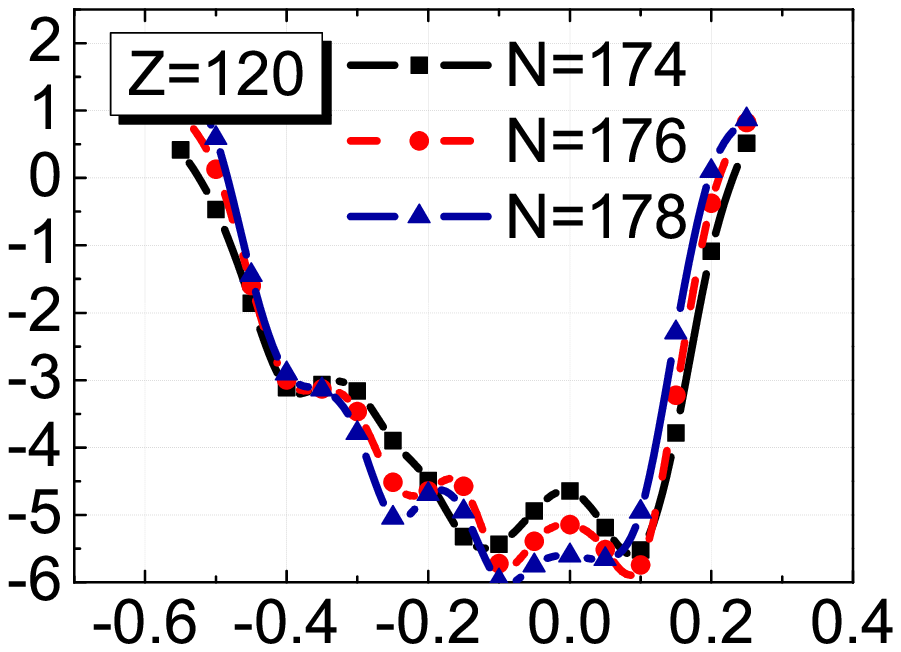}}
\vspace{-7mm}
\centerline{\includegraphics[scale=0.42]{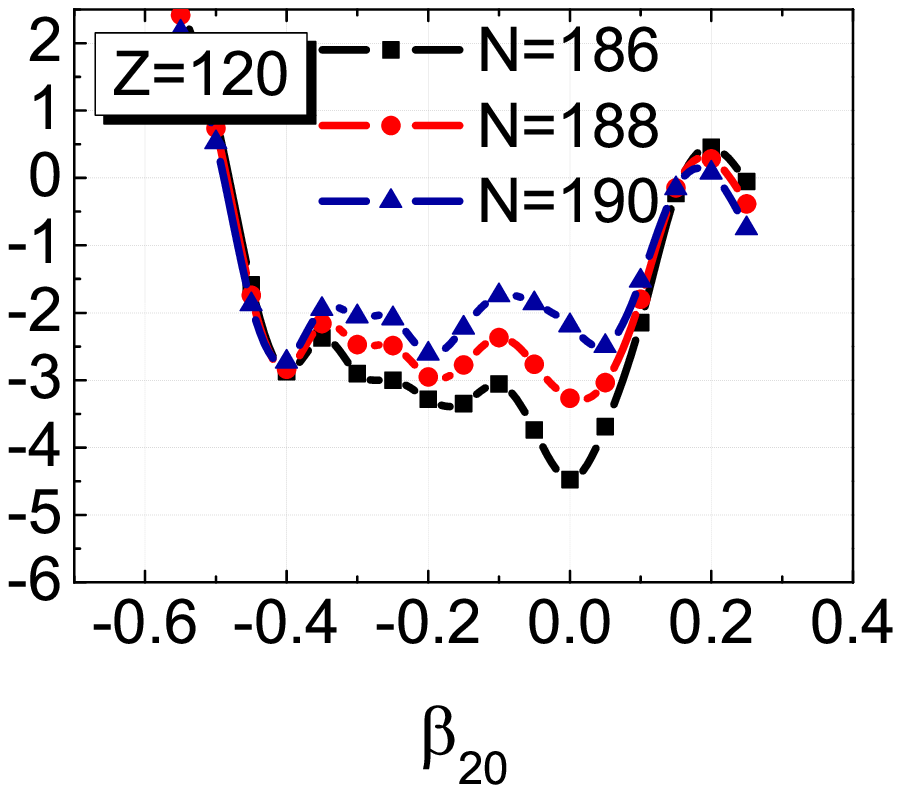}}
\end{minipage}
\caption{{\protect Energy relative to the spherical macroscopic contribution,
    $E(\beta_{20})-E_{macr}$(sphere), for the $Z=120$ isotopic chain; 
   each point results from the minimization over 
   $\beta_{\lambda 0}$, $\lambda=3$-8 (color online).
  }}
 \label{fig:122}
\end{figure}

 {\it Equilibria \& SDO minima; fission barriers}.
  Quadrupole deformations $\beta_{20}$ of the g.s. (global) minima, 
 calculated for $\sim 300$ even-even 
%$ 98\leq Z \leq 126$, $132\leq N \leq 192$ 
 nuclei are shown in Fig. \ref{fig:b2}. 
  In addition to spherical, well- or weakly deformed
  prolate and oblate equilibrium shapes there is a region of
   SDO nuclei for $Z\geq120$, $N\leq168$, of particular interest here. 
 SDO global minima occur also for large $N=190$,192 and $Z=118$-122.  
 Although some weakly deformed minima have non-axial distortions, energies 
 of $Z=120$ isotopes plotted vs. $\beta_{20}$ for axially symmetric shapes  
 in Fig.\ref{fig:122} fairly illustrate the shape competition \& coexistence in 
 the $Z\geq120$ region. The secondary SDO minima exist there for 
 $168\leq N\leq172$ and $N\geq184$. They appear also in $Z\leq 118$ nuclei. 
 Typically, they lie $\approx2$ MeV above the g.s. This has an effect on 
 the $\alpha$-decay of the SDO $Z=120$ isotopes (see below). 
 In the whole $Z\geq114$ region, the deepest minima, spherical or oblate, 
 occur for $N=174$-184; for $Z=124$,126 they are predominantly oblate. 

\begin{figure}[h]
\vspace{-5mm}
\hspace{-3mm}
\begin{minipage}[l]{-4mm}
%\centerline{\includegraphics[scale=0.95]{122_map.PDF}}
\centerline{\includegraphics[scale=0.95]{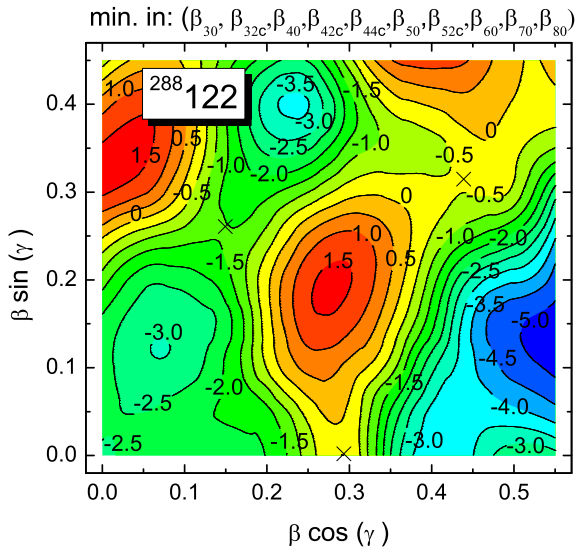}}
\end{minipage}
\caption{{\protect Energy surface of $^{288}{122}$, normalized 
 as in Fig.\ref{fig:122}. Crosses mark the saddles (color online).  
  }}
  \label{fig:Th226}
\end{figure}
   Energy maps in ($\beta \cos\gamma$, $\beta \sin\gamma$) plane are necessary 
   to appreciate fission barriers, Fig \ref{fig:Th226}.   
   The conspicuous result of our calculations is that triaxial saddles
    occur in all studied nuclei. They may lower the axial fission barrier by
 up to 2.5 MeV. This lowering increases with $N$ and is larger for bigger $Z$.  
%  The nonaxial saddle may also provide an alternative fission path, 
%  see Fig. \ref{fig:Th226}. 
    The odd-multipole deformations do not change the barriers as much, but 
    they lower some oblate minima and modify the energy maps  
    around and beyond the saddles.  

   Crucial for stability is that barriers diminish with $N$ {\it decreasing}
  below 174-176 and with $Z$ approaching 126. The first feature is common 
  also to the self-consistent HF results \cite{Sta}, while the second is very 
   distinctive for the macroscopic-microscopic model used here \cite{Kow}.
   Hence, the largest barriers of $\approx 3.4$ MeV predicted for SDO nuclei 
  $^{286}$120 and $^{288}$122 are rather small as compared to the 
  5.6 MeV barrier for $^{296}$120 \cite{Kow}. 
   The barriers for $N\geq 190$ SDO nuclei are still smaller, so we do not 
  consider them further.  
   As the $\alpha$-decay rates increase with $Z$, we concentrate on 
   the SDO nuclei around $Z\approx 120$ and $N\approx 166$.   

 {\it Other models.} To convince ourselves that the SDO minima are not a strange twist of the
  particular model we repeated the minimizations for the interesting nuclei
  by using A) the same microscopic model and another version of the macroscopic
  energy, the LSD liquid drop model of \cite{POM2003}, B) the selfconsistent
  HF method with the realistic Skyrme SLy6 force \cite{SLy6}. 
 % with pairing strength of J. Skalski, Phys. Rev. C  ... 
  Both calculations support the prediction of the global SDO minima;
   they are even by $\approx 1$ MeV deeper with the LSD variant of the 
  macroscopic energy.
  In the HF calculations, the energy competition between 
  prolate, oblate and SDO minima and fission barriers come out similar as 
  in the macroscopic-microscopic study.

 %  The barriers of $\sim $3 MeV seem too low to assure measurable fission half-lives. 
   {\it Stability against fission.}
   We checked fission half-lives $T_{sf}$ by calculating WKB action with 
  cranking mass parameters for selected nuclei. We assumed the zero-point 
  energy of 0.5 MeV.   
   To handle fission paths in 12D deformation space we calculate,  
  instead of the mass parameter tensor, the effective mass parameter along 
 a prescribed path. Technically, this is done by 
  replacing analytic derivatives with respect to deformations  
   by the finite differences.

  Two possible classes of fission paths and barriers along them may be read 
  from Fig. \ref{fig:Th226}. The barriers along the axial saddle 
  (at $\beta\approx0.3$, $\gamma=0$) are longer and have thinner peaks. 
  They can compete with the triaxial path 
   only when there is a deep normal oblate minimum, i.e. for $N=$166 or 168. 
  Triaxial barriers and the related WKB action change smoothly 
  from isotope to isotope. The smallest action we found along 
  triaxial, nearly straight paths. They give half-lives  
  $10^{-6}$ s for $^{286}$120 and $10^{-5}$ s for 
  $^{288}$122, with an estimated error of 1 order of magnitude.   

\begin{figure}[t]
\vspace{-5mm}
\begin{minipage}[l]{2mm}
\centerline{\includegraphics[scale=0.80]{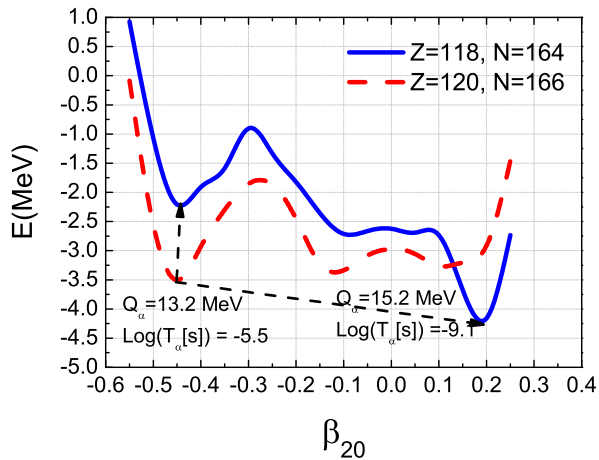}}
\end{minipage}
\caption{{\protect Mechanism of the $\alpha$-decay hindrance of the 
%$^{286}$120 and $^{162}$118
  SDO $^{286}$120; energy normalized as in Fig.\ref{fig:122} 
  (color online).
  }}
  \label{fig:coexistence}
\end{figure}
 {\it Stability against $\alpha$-decay}.
  From the calculated masses and the improved formula a la Viola-Seaborg 
  \cite{Royer}, we obtain for the g.s.$\rightarrow$g.s. transitions  
   $Log(T_{\alpha}[$s$])=-9.1$ for $^{286}$120, and {\it longer} 
 $T_{\alpha}$ for the lighter $Z=120$ isotopes. 
   These SDO$\rightarrow$prolate transitions, Fig.\ref{fig:coexistence},
   must be strongly hindered by a very different structure of both 
   configurations, in particular, the occupation of intruder states at SDO 
   shape (see below). 
  If the hindrance would be complete, only SDO$\rightarrow$SDO 
  transitions would remain. As already mentioned, SDO configurations in 
  the $Z=118$ daughters are excited by $\approx2$ MeV (2.5 MeV in HF). 
   This leads to a considerable increase in half-life: 
   $Log(T_{\alpha}[$s$])$ becomes equal to $-5.5$ for $^{286}$120
  and $T_{\alpha}$ are {\it shorter} for lighter isotopes.
% The additional, smaller hindrance has to be expected due to 
% the systematic difference in deformations, $\Delta \beta_{20}\approx 0.04$,   
% between the SDO configurations in $Z=120$ parents and $Z=118$ daughters. 
   Clearly, this result holds for the configuration hindrance factors  
    $\leq 10^{-3.6}$.                              
  
\begin{figure}[h]
\vspace{-5mm}
\begin{minipage}[h]{45mm}
\centerline{\includegraphics[scale=0.65]{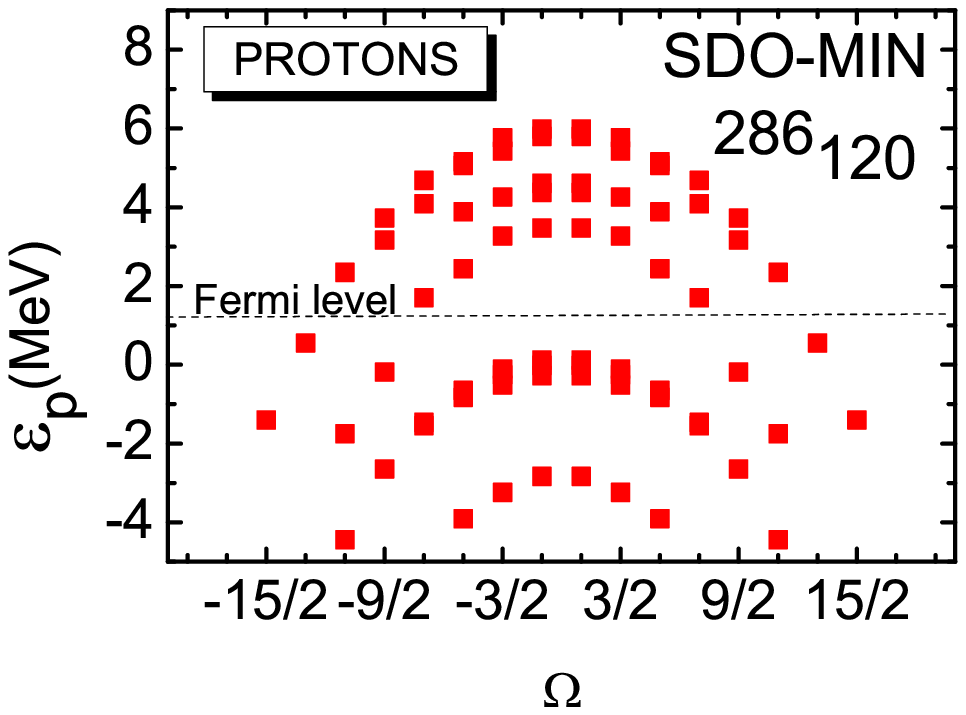}}
\end{minipage}
\vspace{-5mm}
\begin{minipage}[h]{80mm}
\centerline{\includegraphics[scale=0.65]{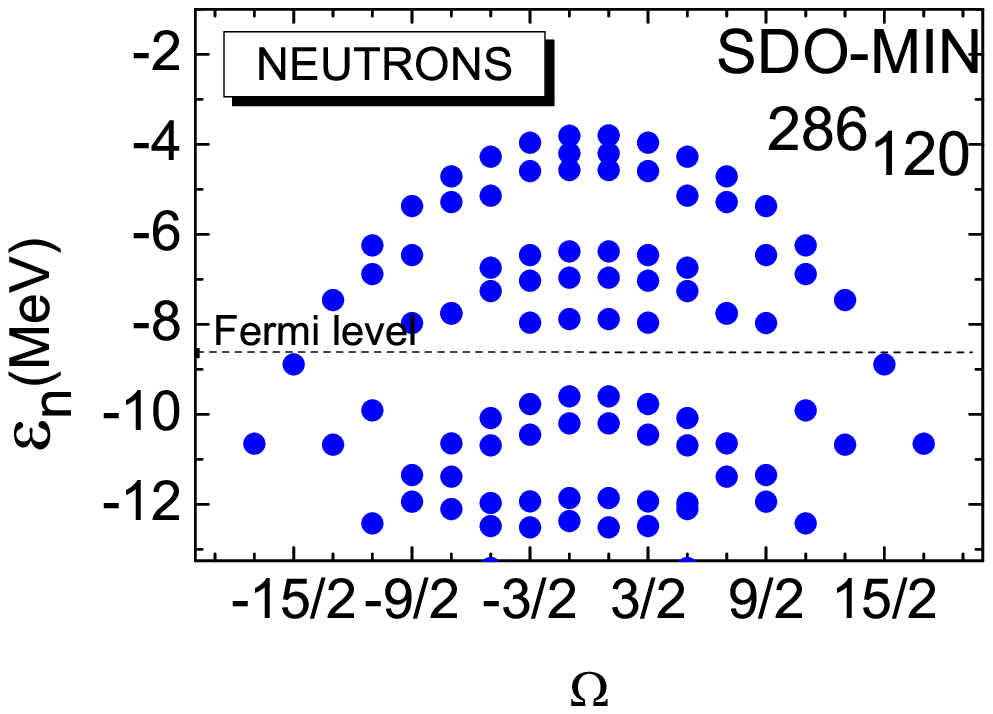}}
\end{minipage}
\caption{{\protect Calculated s.p. energies  vs projection of the angular 
 momentum on the symmetry axis in $^{286}120$ at the SDO g.s.: 
 $\beta_{20}=-0.449$, $\beta_{40}=0.054$, $\beta_{60}=0.017$, 
 $\beta_{80}=-0.017$ (color online).
 }}
 \label{fig:protneut}
\end{figure}

  {\it $K$-isomers at SDO deformation; odd systems}.
   With half-lives $T_{1/2}<10^{-5}$ s - the present limit  
   for detection of synthesized superheavy nuclei - superheavy SDO systems 
   might be considered merely as a theoretical curiosity. 
 A fascinating possibility for their longer life-times is related to $K$-
isomerism, see \cite{marinov,izom}. Indeed, high-$K$ configurations at the 
  SDO shape are very likely, see Fig. \ref{fig:protneut}. 
  Due to large deformation, the neutron $k_{17/2}$ and proton $j_{15/2}$ 
 intruder states with large angular momentum projections on the symmetry 
  axis $\Omega$ are close to the Fermi level for $Z=120$, $N=166$. 
  Of unique structure and parity, they provide identity to high-$K$ 
  2(4)-quasiparticle configurations. Candidates for low-lying $K$-isomers are 
  the so called "optimal" configurations \cite{BM}, with singly occupied 
  large-$\Omega$ orbitals close to the Fermi level. In $^{286}$120, 
  the candidates are the proton (13/2$^-$,7/2$^+$)10$^-$ and neutron 
  (15/2$^+$,9/2$^-$)12$^-$ configurations. The possible low-lying or ground 
  states in odd nuclei are the neutron 15/2$^+$ state in $^{285}$120 and the 
  proton 13/2$^-$ state in $^{285}$119; the low-lying 14$^-$ state 
 could be expected in the odd-odd $^{284}$119. Detailed predictions would 
  require energy minimization at fixed configuration with blocking. 

  In assessing stability of high-$K$ isomers or odd nuclei we  
 rely on estimates and analogies with well established experimental facts,
  as we cannot precisely calculate their decay rates.  
  Let us notice that the considered SDO nuclei are proton-unstable,
  but in view of the large Coulomb barrier the related life-times 
 may not concern us, at least for even-$Z$ nuclei \cite{611}; odd-Z, 
 high-$K$ states are protected by the centrifugal barrier for 
  high-$\Omega$ protons.   

  {\it Fission hindrance}.
  As well known, $T_{sf}$ for odd and odd-odd heavy and superheavy nuclei  
  are by 3-5 orders longer than for their even-even neighbours. 
  Similar increase was found for high-$K$ isomers, with respect to (prolate) 
  shape isomers on which they are built, in even $^{240-244}$Cm \cite{sletten}. 
  For SDO superheavy $K$-isomers two factors combine to increase fission 
  half-life: 
  A) the axial fission path is closed by the conservation of the 
    $K$ quantum number,
 B) triaxial barriers increase due to a decrease in pairing caused by 
   the blocking of two neutrons or protons. Additional hindrance of fission 
   is expected for configurations involving blocked high-$\Omega$ 
   intruder states. 

   Consider now the effect of the saddle deformation on the fission barrier at 
  high spins. The geometrical moments of inertia from the HF calculation in 
  $^{288}$120 are: ${\cal J}_{\perp}=71$ b, ${\cal J}_{\parallel}=109$ b at 
  the SDO shape and ${\cal J}^b_{\perp}=107$ b at the triaxial barrier. 
 The actual moment of inertia at the barrier is reduced by pairing to 
  $f^b {\cal J}^b_{\perp}$, with $f^b$ substantially smaller than 1. 
 Without pairing, ${\cal J}_{\parallel}$ is an {\it average} moment of inertia 
 of yrast non-collective high-$K$ states \cite{BM}. As 
  ${\cal J}_{\parallel}>f^b {\cal J}^b_{\perp}$ and pairing at SDO g.s. 
  is {\it weaker} than at the barrier, there should be {\it no} decrease 
  with $K$ in fission barrier for SDO $K$-isomers. 

 {\it Alpha-decay hindrance}.
  Although this seems the least certain of our arguments, $K$-isomerism 
 {\it may} substantially increase $\alpha$ half-lives:
 the high-$K$ isomer in $^{270}$Ds has longer (partial) half-live
   $T_{\alpha}= 6.0 ^{+8.2}_{-2.2}$ ms than the g.s.,  
  $T_{\alpha}(g.s.)=100 ^{+140}_{-40} $ $\mu$s \cite{Sigurd}.
  For SDO nuclei, an additional hindrance may result from 
  a difference between the parent and daughter high-$K$ configuration, or, 
  for the same configuration, from its extra excitation in the 
    daughter, leading to a smaller $Q_{\alpha}$. 

 {\it Stability against beta-decay}. The $\beta^+$ decay rates 
 $\lambda_{\beta}$ for neutron-deficient candidates for the SDO $K$-isomers  
 can be estimated by neglecting the emitted electron energy m$_e$c$^2$ in the 
  decay energy: $Q_{\beta}=(M(A,Z)-M(A,Z-1) -m_{e})c^2$.
  Then, one has $\lambda_{\beta} \sim |M|^2G_{F}^{2}Q_{\beta}^{5}$, where
  $|M|$ is the transition matrix element and $G_{F}$ is the Fermi constant.
  Even for a perfect overlap, $|M|^2 \sim 1$, using our calculated 
  masses we obtain half-lives $T_{\beta}=ln2/{\lambda_{\beta}}$ 
 of the order of 0.1-1 s for even and odd SDO nuclei, 
 consistent with the results by M\"{o}ller et al. \cite{Mo2}. 
 Since for high-$K$ isomers $|M|$ is reduced, 
 %$^{286}120$
%Even larger half-lives would follow for $\beta$-decays with $|M|^2 \ll 1$. 
  their $\beta^+$ decay is even slower. 

 Although the production of SDO nuclei is another subject, one may notice 
  here that the SDO shape is much closer to the sticking point configuration 
 of the prolate and spherical heavy ions in the side collision 
 than the sphere.   
% 1) The considered shapes have an obvious
%relation to configurations with density depletion in the center -
%a trace of toroidal  shape. 2) 
%  3) The high-$K$ isomers can exist also at the g.s.
%deformation in considered nuclei which could be more stable due to
%higher fission barrier. Since this depends on details of
%s.p. level spectra one cannot at present be sure whether SDO configurations
% would be less stable.

 Summarizing, within both macroscopic-microscopic and Skyrme HF methods,  
 one obtains SDO shapes of the ground- or low excited states of superheavy 
 $Z\geq120$ nuclei. Although even-even, spin zero nuclei decay  
  by a quick $\sim 10^{-5}$-$10^{-6}$ s fission or $\alpha$-decay,  
   longer half-lives are expected for high-$K$ isomers which very likely exist 
   in some even or odd systems. One case of a sizable     
    $\alpha$-decay hindrance could make such a system detectable 
    by the present technique. 

%%%%%%%%%%%%%%%%%%%%%%%%%%%%%%%%%%%%%%%%%%%%%%%%%%%%%%%%%%%%
%\section*{Acknowledgements}
%This work was supported in part by the Polish Ministry of Science and
%  Higher Education, Contract No. N N202 328234.
%
%%%%%%%%%%%%%%%%%%%%%%%%%%%%%%%%%%%%%%%%%%%%%%%%%%%%%%%%%%%%%%%%%%%%%%%kklkk


\begin{thebibliography}{99}
% \bibitem*{114}
%% Yu. Ts. Oganessian et. al, Yad. Fiz. 63 1769 (2000)
%%[Phys. At. Nucl. 63, 1679 (2000)]
% Yu. Ts. Oganessian et al., {\it Phys. Rev. Lett.} {\bf 83} (1999) 3154,
%Yu. Ts. Oganessian et al., {\it Nature} (London) {\bf 400} (1999)
%242, Yu. Ts. Oganessian et al., {\it Phys. Rev. C} {\bf 62} (2000)
%041604, Yu. Ts. Oganessian et al., {\it Phys. Rev. C} {\bf 69}
%(2004) 054607.

% \bibitem{116}
% Yu. Ts. Oganessian et al., {\it Phys. Rev. C} {\bf 63} (2001) 011301(R).
% Yu. Ts. Oganessian et al., {\it Phys. Rev. C} {\bf 63},  011301 (2001).

% \bibitem{115}
% Yu.~Ts.~Oganessian et al., {\it Phys. Rev. C} {\bf 69}, 021601 (2004).

\bibitem{118}
 Yu. Ts. Oganessian et al., {\it Phys. Rev. C} {\bf 74}, 044602 (2006).

\bibitem{GSI}
  TASCA report, GSI Kurier 31 (2009) (unpublished).

% potwierdzenie Dubnej
\bibitem{LBL}
 L.~Stavsetra et al., {\it Phys. Rev. Lett.} {\bf 103}, 132502 (2009).

\bibitem{No}
P. Reiter et al.,
 {\it Phys. Rev. Lett.} {\bf82}, 509–512 (1999).

\bibitem{611}
     S.~\'Cwiok, J. Dobaczewski, P.-H.~Heenen, P.~Magierski and
W.~Nazarewicz,
   {\it Nucl. Phys. A} {\bf 611}, 211 (1996).


\bibitem{CHN}
    S. \'{C}wiok, P.-H. Heenen, W. Nazarewicz, {\it Nature} {\bf433}, 709
(2005).

\bibitem{marinov}
 A.~Marinov, S.~Gelberg, D.~Kolb and J.~L. ~Weil, {\it Int. J.
Mod. Phys.} E10, 3 (2001).

 \bibitem{WS}
    S.~\'Cwiok, J.~Dudek, W.~Nazarewicz, J.~Skalski and T.~Werner,
 {\it Comput. Phys. Commun.} {\bf 46}, 379 (1987).


\bibitem{KN}
  H.~J.~Krappe, J.~R.~Nix and A.~J.~Sierk, {\it Phys. Rev. C} {\bf 20},
992 (1979).

\bibitem{WSpar}
  I.~Muntian, Z.~Patyk and A.~Sobiczewski, {\it Acta Phys. Pol. B} {\bf
  32}, 691 (2001).

\bibitem{Kow}
 M. Kowal, P. Jachimowicz, A. Sobiczewski, {\it Phys. Rev. C} {\bf 82},
014303 (2010).


\bibitem{Sta}
  A.~Staszczak, J.~Dobaczewski and W.~Nazarewicz, {\it Int. J. Mod. Phys. E}
 {\bf 15}, 302 (2006).

% \bibitem{KR}
%  K.~Rykaczewski, private communication.


 \bibitem{POM2003}
 K. Pomorski, J. Dudek,  {\it Phys. Rev. C} {\bf67}, 044316 (2003).


\bibitem{SLy6}
E. Chabanat et al., {\it Nucl. Phys.} {\bf A635} (1998) 231.

 \bibitem{Royer}
G. Royer,  K. Zbiri and C. Bonilla,
  { \it Nuclear Physics A} 730  (2004).

 \bibitem{izom}
 F.~R.~Xu, E.~G.~Zhao, R.~Wyss and P.~M.~Walker, {\it Phys. Rev. Lett.}
{\bf 92}, 252501 (2004).

 \bibitem{BM}
 A.~Bohr, B.~R.~Mottelson, {\it Nuclear Structure} Vol. 2 (Benjamin,New York,1975) 
 \bibitem{sletten}
 H.~C.~Britt, S.~C.~Burnett, B.~H.~Erkkila, J.~E.~Lynn 
 and W.~E.~Stein, {\it Phys. Rev. C} {\bf 4}, 1444 (1971); G.~Sletten, 
 V.~Metag and E.~Liukkonen, {\it Physics Letters B} {\bf 60}, 2 (1976).

 \bibitem{Sigurd}
    S. Hofmann et al., {\it Eur. Phys. J. A} {\bf10}, 5 (2001).



% \bibitem{Mo}
% P. M\"oller et al., {\it Phys. Rev. C} {\bf 79}, 064304 (2009).

  % rozp. beta
\bibitem{Mo2}
 P. M\"oller et al., { \it Atomic Data and Nuclear Data Tables} {\bf 66}
2, 131 (1997).


%\bibitem{bjor}
%S.~Bj\o rnholm, J.~E. Lynn, {\it Rev. Mod. Phys.} {\bf 52}, 725
%(1980) and references therein.

%\bibitem{sign}
%B.~Singh, R.~Zywina, R.~B. ~Firestone,  Table of Superdeformed
%Nuclear Bands and Fission Isomers, {\it Nucl. Data Sheets} {\bf
%97}, 241 (2002).




\end{thebibliography}
\end{document}